\documentclass[longauth]{aa}

\usepackage{graphicx}
\graphicspath{{figures/}}
\usepackage{txfonts}
\usepackage[colorlinks, allcolors=blue]{hyperref}
            
\usepackage{euclid} 
\usepackage{multirow} 
\usepackage[nameinlink]{cleveref} 

\crefname{section}{Sect.}{Sects.}
\crefname{figure}{Fig.}{Figs.}
\crefname{equation}{Eq.}{Eqs.}
\crefname{table}{Table}{Tables}

\begin{document}

\def\equationautorefname{Eq.} 

\title{\Euclid: Covariance of weak lensing pseudo-$C_\ell$ estimates. Calculation, comparison to simulations, and dependence on survey geometry\thanks{This paper is published on behalf of the Euclid Consortium.}}
\titlerunning{\Euclid: Covariance of weak lensing pseudo-$C_\ell$ estimates}

\newcommand{\lastleadauthor}[0]{L.~Whittaker} 
\author{R.E.~Upham$^{1}$\thanks{\email{robin.upham@manchester.ac.uk}},
M.L.~Brown$^{1}$,
L.~Whittaker$^{2,1}$,
A.~Amara$^{3}$,
N.~Auricchio$^{4}$,
D.~Bonino$^{5}$,
E.~Branchini$^{6,7}$,
M.~Brescia$^{8}$,
J.~Brinchmann$^{9}$,
V.~Capobianco$^{5}$
% ...
}
\authorrunning{R.E.~Upham et al.}

\institute{$^{1}$ Jodrell Bank Centre for Astrophysics, Department of Physics and Astronomy, University of Manchester, Oxford Road, Manchester M13 9PL, UK\\
$^{2}$ Department of Physics and Astronomy, University College London, Gower Street, London WC1E 6BT, UK\\
$^{3}$ Institute of Cosmology and Gravitation, University of Portsmouth, Portsmouth PO1 3FX, UK\\
$^{4}$ INAF-Osservatorio di Astrofisica e Scienza dello Spazio di Bologna, Via Piero Gobetti 93/3, I-40129 Bologna, Italy\\
$^{5}$ INAF-Osservatorio Astrofisico di Torino, Via Osservatorio 20, I-10025 Pino Torinese (TO), Italy\\
$^{6}$ Department of Mathematics and Physics, Roma Tre University, Via della Vasca Navale 84, I-00146 Rome, Italy\\
$^{7}$ INFN-Sezione di Roma Tre, Via della Vasca Navale 84, I-00146, Roma, Italy\\
$^{8}$ INAF-Osservatorio Astronomico di Capodimonte, Via Moiariello 16, I-80131 Napoli, Italy\\
$^{9}$ Instituto de Astrof\'isica e Ci\^encias do Espa\c{c}o, Universidade do Porto, CAUP, Rua das Estrelas, PT4150-762 Porto, Portugal\\
$^{10}$ INAF-IASF Milano, Via Alfonso Corti 12, I-20133 Milano, Italy\\
$^{11}$ Port d'Informaci\'{o} Cient\'{i}fica, Campus UAB, C. Albareda s/n, 08193 Bellaterra (Barcelona), Spain\\
$^{12}$ Institut de F\'{i}sica d’Altes Energies (IFAE), The Barcelona Institute of Science and Technology, Campus UAB, 08193 Bellaterra (Barcelona), Spain\\
$^{13}$ INAF-Osservatorio Astronomico di Roma, Via Frascati 33, I-00078 Monteporzio Catone, Italy\\
$^{14}$ Department of Physics "E. Pancini", University Federico II, Via Cinthia 6, I-80126, Napoli, Italy\\
$^{15}$ INFN section of Naples, Via Cinthia 6, I-80126, Napoli, Italy\\
$^{16}$ Dipartimento di Fisica e Astronomia “Augusto Righi” - Alma Mater Studiorum Università di Bologna, via Piero Gobetti 93/2, I-40129 Bologna, Italy\\
$^{17}$ INAF-Osservatorio Astrofisico di Arcetri, Largo E. Fermi 5, I-50125, Firenze, Italy\\
$^{18}$ Centre National d'Etudes Spatiales, Toulouse, France\\
$^{19}$ Institut national de physique nucl\'eaire et de physique des particules, 3 rue Michel-Ange, 75794 Paris C\'edex 16, France\\
$^{20}$ Institute for Astronomy, University of Edinburgh, Royal Observatory, Blackford Hill, Edinburgh EH9 3HJ, UK\\
$^{21}$ ESAC/ESA, Camino Bajo del Castillo, s/n., Urb. Villafranca del Castillo, 28692 Villanueva de la Ca\~nada, Madrid, Spain\\
$^{22}$ European Space Agency/ESRIN, Largo Galileo Galilei 1, 00044 Frascati, Roma, Italy\\
$^{23}$ Univ Lyon, Univ Claude Bernard Lyon 1, CNRS/IN2P3, IP2I Lyon, UMR 5822, F-69622, Villeurbanne, France\\
$^{24}$ Mullard Space Science Laboratory, University College London, Holmbury St Mary, Dorking, Surrey RH5 6NT, UK\\
$^{25}$ Departamento de F\'isica, Faculdade de Ci\^encias, Universidade de Lisboa, Edif\'icio C8, Campo Grande, PT1749-016 Lisboa, Portugal\\
$^{26}$ Instituto de Astrof\'isica e Ci\^encias do Espa\c{c}o, Faculdade de Ci\^encias, Universidade de Lisboa, Campo Grande, PT-1749-016 Lisboa, Portugal\\
$^{27}$ Department of Astronomy, University of Geneva, ch. d\'Ecogia 16, CH-1290 Versoix, Switzerland\\
$^{28}$ Universit\'e Paris-Saclay, CNRS, Institut d'astrophysique spatiale, 91405, Orsay, France\\
$^{29}$ Department of Physics, Oxford University, Keble Road, Oxford OX1 3RH, UK\\
$^{30}$ INFN-Padova, Via Marzolo 8, I-35131 Padova, Italy\\
$^{31}$ AIM, CEA, CNRS, Universit\'{e} Paris-Saclay, Universit\'{e} de Paris, F-91191 Gif-sur-Yvette, France\\
$^{32}$ Institut d’Estudis Espacials de Catalunya (IEEC), Carrer Gran Capit\'a 2-4, 08034 Barcelona, Spain\\
$^{33}$ Institute of Space Sciences (ICE, CSIC), Campus UAB, Carrer de Can Magrans, s/n, 08193 Barcelona, Spain\\
$^{34}$ INAF-Osservatorio Astronomico di Trieste, Via G. B. Tiepolo 11, I-34131 Trieste, Italy\\
$^{35}$ Istituto Nazionale di Astrofisica (INAF) - Osservatorio di Astrofisica e Scienza dello Spazio (OAS), Via Gobetti 93/3, I-40127 Bologna, Italy\\
$^{36}$ Istituto Nazionale di Fisica Nucleare, Sezione di Bologna, Via Irnerio 46, I-40126 Bologna, Italy\\
$^{37}$ Max Planck Institute for Extraterrestrial Physics, Giessenbachstr. 1, D-85748 Garching, Germany\\
$^{38}$ Universit\"ats-Sternwarte M\"unchen, Fakult\"at f\"ur Physik, Ludwig-Maximilians-Universit\"at M\"unchen, Scheinerstrasse 1, 81679 M\"unchen, Germany\\
$^{39}$ Institute of Theoretical Astrophysics, University of Oslo, P.O. Box 1029 Blindern, N-0315 Oslo, Norway\\
$^{40}$ Leiden Observatory, Leiden University, Niels Bohrweg 2, 2333 CA Leiden, The Netherlands\\
$^{41}$ Jet Propulsion Laboratory, California Institute of Technology, 4800 Oak Grove Drive, Pasadena, CA, 91109, USA\\
$^{42}$ von Hoerner \& Sulger GmbH, Schlo{\ss}Platz 8, D-68723 Schwetzingen, Germany\\
$^{43}$ Technical University of Denmark, Elektrovej 327, 2800 Kgs. Lyngby, Denmark\\
$^{44}$ Max-Planck-Institut f\"ur Astronomie, K\"onigstuhl 17, D-69117 Heidelberg, Germany\\
$^{45}$ Aix-Marseille Univ, CNRS/IN2P3, CPPM, Marseille, France\\
$^{46}$ Universit\'e de Gen\`eve, D\'epartement de Physique Th\'eorique and Centre for Astroparticle Physics, 24 quai Ernest-Ansermet, CH-1211 Gen\`eve 4, Switzerland\\
$^{47}$ Department of Physics and Helsinki Institute of Physics, Gustaf H\"allstr\"omin katu 2, 00014 University of Helsinki, Finland\\
$^{48}$ NOVA optical infrared instrumentation group at ASTRON, Oude Hoogeveensedijk 4, 7991PD, Dwingeloo, The Netherlands\\
$^{49}$ Argelander-Institut f\"ur Astronomie, Universit\"at Bonn, Auf dem H\"ugel 71, 53121 Bonn, Germany\\
$^{50}$ INFN-Sezione di Bologna, Viale Berti Pichat 6/2, I-40127 Bologna, Italy\\
$^{51}$ Institute of Physics, Laboratory of Astrophysics, Ecole Polytechnique F\'{e}d\'{e}rale de Lausanne (EPFL), Observatoire de Sauverny, 1290 Versoix, Switzerland\\
$^{52}$ European Space Agency/ESTEC, Keplerlaan 1, 2201 AZ Noordwijk, The Netherlands\\
$^{53}$ Department of Physics and Astronomy, University of Aarhus, Ny Munkegade 120, DK–8000 Aarhus C, Denmark\\
$^{54}$ Institute of Space Science, Bucharest, Ro-077125, Romania\\
$^{55}$ IFPU, Institute for Fundamental Physics of the Universe, via Beirut 2, 34151 Trieste, Italy\\
$^{56}$ Dipartimento di Fisica e Astronomia “G.Galilei", Universit\'a di Padova, Via Marzolo 8, I-35131 Padova, Italy\\
$^{57}$ Centro de Investigaciones Energ\'eticas, Medioambientales y Tecnol\'ogicas (CIEMAT), Avenida Complutense 40, 28040 Madrid, Spain\\
$^{58}$ Instituto de Astrof\'isica e Ci\^encias do Espa\c{c}o, Faculdade de Ci\^encias, Universidade de Lisboa, Tapada da Ajuda, PT-1349-018 Lisboa, Portugal\\
$^{59}$ Universidad Polit\'ecnica de Cartagena, Departamento de Electr\'onica y Tecnolog\'ia de Computadoras, 30202 Cartagena, Spain\\
$^{60}$ Infrared Processing and Analysis Center, California Institute of Technology, Pasadena, CA 91125, USA\\
$^{61}$ INAF-Osservatorio Astronomico di Brera, Via Brera 28, I-20122 Milano, Italy\\
$^{62}$ Dipartimento di Fisica e Astronomia, Universit\'a di Bologna, Via Gobetti 93/2, I-40129 Bologna, Italy\\
$^{63}$ Dipartimento di Fisica, Universit\'a degli Studi di Torino, Via P. Giuria 1, I-10125 Torino, Italy\\
$^{64}$ INFN-Sezione di Torino, Via P. Giuria 1, I-10125 Torino, Italy\\
$^{65}$ INFN-Sezione di Roma, Piazzale Aldo Moro, 2 - c/o Dipartimento di Fisica, Edificio G. Marconi, I-00185 Roma, Italy\\
$^{66}$ Center for Computational Astrophysics, Flatiron Institute, 162 5th Avenue, 10010, New York, NY, USA\\
$^{67}$ School of Physics and Astronomy, Cardiff University, The Parade, Cardiff, CF24 3AA, UK\\
$^{68}$ Space Science Data Center, Italian Space Agency, via del Politecnico snc, 00133 Roma, Italy\\
$^{69}$ Astrophysics Group, Blackett Laboratory, Imperial College London, London SW7 2AZ, UK\\
$^{70}$ Mathematical Institute, University of Leiden, Niels Bohrweg 1, 2333 CA Leiden, The Netherlands\\
}

\abstract{An accurate covariance matrix is essential for obtaining reliable cosmological results when using a Gaussian likelihood. In this paper we study the covariance of pseudo-$C_\ell$ estimates of tomographic cosmic shear power spectra. Using two existing publicly available codes in combination, we calculate the full covariance matrix, including mode-coupling contributions arising from both partial sky coverage and non-linear structure growth. For three different sky masks, we compare the theoretical covariance matrix to that estimated from publicly available N-body weak lensing simulations, finding good agreement. 
We find that as a more extreme sky cut is applied, a corresponding increase in both Gaussian off-diagonal covariance and non-Gaussian super-sample covariance is observed in both theory and simulations, in accordance with expectations.
Studying the different contributions to the covariance in detail, we find that the Gaussian covariance dominates along the main diagonal and the closest off-diagonals, but farther away from the main diagonal the super-sample covariance is dominant.
Forming mock constraints in parameters that describe matter clustering and dark energy, we find that neglecting non-Gaussian contributions to the covariance can lead to underestimating the true size of confidence regions by up to 70 per cent. The dominant non-Gaussian covariance component is the super-sample covariance, but neglecting the smaller connected non-Gaussian covariance can still lead to the underestimation of uncertainties by 10--20 per cent. A real cosmological analysis will require marginalisation over many nuisance parameters, which will decrease the relative importance of all cosmological contributions to the covariance, so these values should be taken as upper limits on the importance of each component.}

\keywords{gravitational lensing: weak -- methods: statistical -- cosmology: observations}
                
\maketitle

\section{Introduction}

There are currently many unanswered questions in cosmology, including the origin of the accelerating expansion of the universe and apparent tensions within the dominant $\Lambda$ cold dark matter model \citep[e.g.][and references therein]{DiValentino2021}. One of the most promising tools with which to make progress on these questions in the coming years is the analysis of weak gravitational lensing of distant galaxies by large-scale structure, also known as cosmic shear.
The upcoming ESA \Euclid space mission\footnote{\href{https://www.euclid-ec.org}{https://www.euclid-ec.org}} \citep{Laureijs2011}, as well as other surveys such as those with the Vera C. Rubin Observatory in Chile (the Legacy Survey of Space and Time\footnote{\href{https://www.lsst.org}{https://www.lsst.org}}; \citealt{Ivezic2019}) and the Square Kilometre Array (SKA) radio observatory in Australia and South Africa\footnote{\href{https://www.skatelescope.org}{https://www.skatelescope.org}} (\citealt{SKA2020}),
will observe over a billion galaxies, which is expected to lead to unprecedented precision on cosmological constraints -- a more than an order of magnitude increase over the previous generation of experiments \citep{Harrison2016}. In order to obtain reliable results, this precision is necessarily accompanied by a requirement to understand all elements of an analysis pipeline to an equally unprecedented degree, including the interplay between the likelihood and estimator effects. 

In this paper we are concerned specifically with pseudo-$C_\ell$ estimators, which have been used previously for the analysis of weak lensing data from the Hyper-Suprime Cam (HSC) Subaru Strategic Program\footnote{\href{https://hsc.mtk.nao.ac.jp}{https://hsc.mtk.nao.ac.jp}} (\citealt{Miyazaki2012}) in \citet{Hikage2019} and the Dark Energy Survey\footnote{\href{https://www.darkenergysurvey.org}{https://www.darkenergysurvey.org}} (DES; \citealt{DES2005}) in \citet{Camacho2021} and will be used in the analysis of future \Euclid data \citep{Loureiro2021}. It was recently shown in \citet{Upham2021} that a Gaussian likelihood is sufficient to obtain accurate cosmological results from weak lensing pseudo-$C_\ell$ estimates. An important ingredient for a Gaussian likelihood is the covariance matrix, so in this paper we focus on the calculation of a cosmic shear pseudo-$C_\ell$ covariance matrix.

The problem of calculating covariance matrices for weak lensing has been extensively discussed in the literature, ranging from analytic or semi-analytic approaches \citep{Cooray2001, Schneider2002, Joachimi2008cov, Takada2009, Pielorz2010, Hilbert2011, Barreira2018ssc, Hall2019, GouyouBeauchamps2021} through to estimation from simulations \citep{Sato2011, Harnois-Deraps2015, Sellentin2016, Sellentin2016b, Harnois-Deraps2018, Harnois-Deraps2019, Sgier2019, Schneider2020}. Here we extend this work to focus specifically on the covariance of pseudo-$C_\ell$ estimates, for which coupling between modes occurs due to the effect of incomplete sky coverage. This effect is in addition to the non-Gaussian mode coupling that is inherent in weak lensing data as a result of non-linear structure growth, which is known to be important for parameter inference \citep{Sato2013, Barreira2018b}.

In \cref{sec:theory} we discuss the different Gaussian and non-Gaussian components of the cosmic shear pseudo-$C_\ell$ covariance and their implementation in existing publicly available code. We compare this theoretical covariance to that measured from publicly available weak lensing simulations in \cref{sec:sims}. In \cref{sec:importance} we examine the relative importance of the different covariance contributions and how this depends on the mask, which describes the details of sky coverage. 
This part of our analysis shares some similarities with that of \citet{Barreira2018b}, who also studied the relative importance of the different contributions to the cosmic shear covariance for a \Euclid{}-like survey and concluded that the `connected non-Gaussian' component (see \cref{sec:theory}) can be neglected for only a $\lesssim$ 5 per cent underestimation in single-parameter 1$\sigma$ errors. However, in this paper we specifically study pseudo-$C_\ell$ estimates, for which the survey mask mixes power between all multipoles and induces correlations even for Gaussian fields, which for many covariance elements dominate over other sources of correlation (see \cref{sec:sims}). This effect was not included in the analysis of \citet{Barreira2018b}, who assumed a diagonal Gaussian covariance, and its inclusion may lead to different conclusions about the relative importance of the different contributions to the covariance.
We discuss our conclusions in \cref{sec:conclusions}.

\section{Cosmic shear power spectrum covariance contributions}
\label{sec:theory}

We begin this section by summarising the different contributions to the cosmic shear power spectrum covariance. We refer the reader to \citet{Barreira2018ssc} and the other references provided both therein and below for a thorough theoretical background and derivation.

Starting in three-dimensional space, the covariance of the matter power spectrum receives two contributions: one that depends on the matter power spectrum itself, and one that depends on a particular (`parallelogram') configuration of the matter trispectrum that corresponds to the Fourier transform of the connected four-point correlation function. For a Gaussian matter distribution, only the first contribution is non-vanishing, and hence it is commonly referred to as the `Gaussian covariance', which we will use here. (It is also sometimes referred to as the `disconnected' covariance.) We follow \citet{Barreira2018ssc} in referring to the second contribution as the `connected non-Gaussian' component.

However, for any realistic finite-volume survey such as \Euclid, the observed matter power spectrum is convolved with a three-dimensional window function. While the Gaussian and non-Gaussian terms remain distinct, this has the effect of introducing additional non-Gaussian couplings between large-scale modes outside the survey and small-scale modes within the survey. This is commonly known as `super-sample' (originally `beat-coupling'; \citealt{Hamilton2006}) covariance, and physically can be explained by the fact that unobservable large-scale modes within which the survey is embedded can influence the rate of small-scale non-linear structure growth, and therefore also the strength of coupling between small-scale modes \citep{Takada2013, Barreira2018ssc}. Perhaps counter-intuitively, it turns out that this is generally the dominant source of non-Gaussian covariance \citep{Hamilton2006, Barreira2018b}.

Progressing to projected two-point statistics such as cosmic shear angular power spectra, the same three components -- Gaussian, super-sample and connected non-Gaussian -- contribute to the covariance. Strictly speaking, the separation of the super-sample and connected non-Gaussian components is only exact under the Limber approximation \citep{Barreira2018ssc}, but the inaccuracy of the Limber approximation is only relevant on very large scales (very low multipoles, $\ell \lesssim 20$) where non-Gaussian correlations are small.

We will now discuss the calculation of each of the three cosmic shear covariance components in turn.

\subsection{Gaussian covariance}

To calculate the Gaussian covariance, we used the `improved narrow kernel approximation' method \citep{Nicola2021} implemented using the publicly available code \texttt{NaMaster}\footnote{\href{https://github.com/LSSTDESC/NaMaster}{https://github.com/LSSTDESC/NaMaster}} \citep{Alonso2019,Garcia-Garcia2019}. We now provide further details and some background on this method.

The Gaussian covariance component of a general statistically isotropic field on the sphere is equivalent to the total covariance of a Gaussian field with the same power spectrum. The analytic covariance of pseudo-$C_\ell$ estimates on Gaussian fields has been well studied in the cosmic microwave background literature \citep[for a non-Gaussian likelihood method]{Efstathiou2004, Challinor2005, Brown2005, Upham2020} as well as in the context of weak lensing \citep{Garcia-Garcia2019, Nicola2021}. The exact Gaussian pseudo-$C_\ell$ covariance can be written down analytically, and includes terms of the following form \citep[e.g.][]{Brown2005}:
\begin{equation}
\begin{aligned}
\text{Cov} \Big( \widetilde{C}_\ell,~ \widetilde{C}_{\ell'} \Big) = 
&\sum_{\substack{m,~m'\\\ell_1,~\ell_2\\m_1,~m_2}}
W_{\ell \ell_1}^{m m_1} \left( W_{\ell' \ell_1}^{m' m_1} \right)^*
W_{\ell' \ell_2}^{m' m_2} \left( W_{\ell \ell_2}^{m m_2} \right)^* 
C_{\ell_1} C_{\ell_2} \\
&+ \text{similar terms},
\end{aligned}
\label{eqn:pcl_cov}
\end{equation}
where the harmonic space window functions $W$ are given in Eq.~8 of \cite{Brown2005}, and the `similar terms' involve different combinations of power spectra depending on the situation and spins being considered \citep{Hansen2003, Challinor2005}.

The evaluation of \cref{eqn:pcl_cov} requires $\mathcal{O}(\ell_\text{max}^6)$ operations per term, so it is impractical to evaluate exactly and in practice approximations are used. These commonly involve substitutions of the following kind \citep{Efstathiou2004,Brown2005,Garcia-Garcia2019}:
\begin{equation}
C_{\ell_1} C_{\ell_2} \rightarrow C_\ell C_{\ell'},
\label{eqn:nka}
\end{equation}
which allows the power spectrum dependence to be brought out of the sums in \cref{eqn:pcl_cov}. This means that the coefficients in the similar terms are now all the same (except for any possible spin dependence, or if different fields use different masks). Symmetry properties of the harmonic space window function allow the calculation of these coefficients to be further simplified, to the point where the covariance can be evaluated in a reasonable time. In essence, the approximation in \cref{eqn:nka} assumes that the power spectrum is constant over the region around a given $\ell$ in which the window function is non-negligible. This will be accurate as long as the window function is sufficiently sharply peaked, and therefore this approximation is often known as the `narrow kernel approximation' (NKA).
A generalised version of the NKA is described in \cite{Garcia-Garcia2019} and implemented in \texttt{NaMaster}, which supports an arbitrary number of correlated spin-0 and spin-2 fields and has both curved-sky and flat-sky support. We used the curved-sky spin-2 version, which naturally accounts for $E$--$B$ leakage (we assume noise-only $B$-modes). By default, \texttt{NaMaster} provides the covariance of deconvolved pseudo-$C_\ell$ estimates, but we set the \texttt{coupled=True} option to instead obtain the covariance of `raw', un-deconvolved estimates such as those produced by the pseudo-$C_\ell$ estimator developed for \Euclid{}, which is described in \citet{Loureiro2021}. We did not apply any $E$--$B$ purification or noise de-biasing either.

\citet{Nicola2021} introduced a small modification to the NKA that turns out to significantly increase its accuracy, which they refer to as the improved NKA. It involves simply replacing each $C_\ell$ in the standard NKA by its mode-coupled counterpart,
\begin{equation}
C_\ell \rightarrow 
\frac{\langle \widetilde{C}_\ell \rangle}{f_\text{sky}}
= \frac{\sum_{\ell'} \tens{M}_{\ell \ell'} C_{\ell'}}{f_\text{sky}},
\end{equation}
where $\tens{M}$ is the usual pseudo-$C_\ell$ mixing or mode-coupling matrix (see \citealt{Brown2005} for a full derivation of its calculation), and the division by the sky fraction $f_\text{sky}$ is to avoid double-counting the loss of power on the cut sky. \texttt{NaMaster} includes the functionality to calculate the mixing matrix and to apply it to power spectra to calculate $\langle \widetilde{C}_\ell \rangle$, so the extension to the improved NKA is trivial. The recent pseudo-$C_\ell$ analysis of DES Year 1 observations in \citet{Camacho2021} provided the first application of the improved NKA to real data, but takes the approach of deconvolution and noise subtraction to obtain unbiased estimates of the underlying power spectrum, unlike the forward-modelling approach taken in this paper.

\subsection{Super-sample covariance}
\label{sec:ssc}

To calculate the super-sample covariance contribution we used the publicly available code \texttt{CosmoLike}\footnote{\href{https://github.com/CosmoLike}{https://github.com/CosmoLike}} \citep{Krause2017CosmoLike}. Specifically, we used an adapted version of the \texttt{CosmoCov}\footnote{\href{https://github.com/CosmoLike/CosmoCov}{https://github.com/CosmoLike/CosmoCov}} correlation function covariance package \citep{Fang2020CosmoCov}, which obtains the real-space non-Gaussian covariance as a transform of the harmonic space covariance. \texttt{CosmoCov} has been used for the DES Year 1 and Year 3 cosmological analyses \citep{Krause2017, Krause2021, Friedrich2021}, as well as for the non-Gaussian covariance in the Year 1 pseudo-$C_\ell$ analysis in \citet{Camacho2021}. Our adapted code to expose the harmonic space covariance directly is available at \href{https://github.com/robinupham/CosmoCov_ClCov}{https://github.com/robinupham/CosmoCov\_ClCov}.

\texttt{CosmoLike} calculates the super-sample covariance using the approach introduced in \cite{Takada2013}, by considering the response of the small-scale non-linear matter power spectrum to changes in the mean linear density field \citep[see also][]{Chiang2014,Li2014}. The response of the non-linear matter power spectrum is evaluated using a halo model, with the details given in Eqs. A7--A11 of \cite{Krause2017CosmoLike}. We note that since that paper was published, the calculation of the survey variance $\sigma_b (z)$ in their Eq. A8 has been replaced within \texttt{CosmoLike} with the following calculation:
\begin{equation}
\sigma_b \left( z \right) = \frac{1}{4 \pi r^2} 
\frac{1}{C_{\ell = 0}^\text{mask}}
\sum_{\ell = 0}^{1000}
\left( 2 \ell + 1 \right) \, C_\ell^\text{mask} \,
P_\text{lin} \left( \frac{\ell + 1/2}{r}, \, z \right),
\label{eqn:survey_variance}
\end{equation}
where $r = f_\kappa \left( \chi \left( z \right) \right)$, and $C_\ell^\text{mask}$ is the power spectrum of the mask.
This treatment of the cut sky using the mask power spectrum was derived in \citet{Barreira2018ssc}.
The value of $\ell_\text{max} = 1000$ in \cref{eqn:survey_variance} is arbitrary, but has a negligible impact in practice because the power spectrum of both masks is negligible above $\ell \sim 100$.
We note that \texttt{CosmoLike} also provides an alternative implementation of both the super-sample and connected non-Gaussian covariance using the `response approach', which additionally accounts for a tidal contribution to the super-sample covariance and has been found to be more accurate than the standard implementation \citep{Wagner2015, Barreira2017a, Barreira2017b, Barreira2018ssc, Schmidt2018}. Here we use the standard \texttt{CosmoLike} implementation as used by the DES Collaboration \citep{Krause2017, Krause2021, Friedrich2021, Camacho2021}.

As stated previously, we wish to forward-model the effect of the mask to obtain the covariance of raw un-deconvolved pseudo-$C_\ell$ estimates following \citet{Loureiro2021}. 
The non-Gaussian covariance output from \texttt{CosmoLike} corresponds to the covariance of unbiased estimates of the underlying power spectrum $\widehat{C}_\ell$, and does not account for any estimator effects such as cut-sky mode coupling. Within the context of the pseudo-$C_\ell$ method, the closest way to interpret the covariance output from \texttt{CosmoLike} is as the covariance of deconvolved pseudo-$C_\ell$ estimates. This is how it is interpreted in \citet{Camacho2021} and how we chose to interpret it here. In this case, the unbiased estimates $\widehat{C}_\ell$ may in principle be obtained from raw pseudo-$C_\ell$ estimates $\widetilde{C}_\ell$ as
\begin{equation}
\widehat{C}_\ell = \sum_{\ell'} \tens{M}^{-1}_{\ell \ell'} \widetilde{C}_{\ell'}.
\end{equation}
It follows that the covariance matrices of $\widehat{C}_\ell$ and $\widetilde{C}_\ell$, which we denote respectively as $\mathbf{\widehat{\Sigma}}$ and $\mathbf{\widetilde{\Sigma}}$, are related as
\begin{equation}
\mathbf{\widehat{\Sigma}} = 
\left( \tens{M}^{-1} \right) 
\mathbf{\widetilde{\Sigma}} 
\left( \tens{M}^{-1} \right)^\intercal.
\label{eqn:cov_transform_inverse}
\end{equation}
We interpret the covariance as output from \texttt{CosmoLike} as $\mathbf{\widehat{\Sigma}}$. This choice is necessarily an approximation since \texttt{CosmoLike} does not account for any estimator effects, including pseudo-$C_\ell$ mode coupling, but it is a necessary choice and is equivalent to that made in the DES Y1 pseudo-$C_\ell$ analysis in \citet{Camacho2021}. In both that paper and this one (\cref{sec:sims}), the resulting covariance is compared to simulations, with good agreement. A full general non-Gaussian pseudo-$C_\ell$ covariance is presented in \citet{Shirasaki2015}, but in practice an approximation such as the one we make here is necessary.
We therefore inverted \cref{eqn:cov_transform_inverse} to obtain the relation that we used to transform $\mathbf{\widehat{\Sigma}}$ to the raw pseudo-$C_\ell$ covariance, $\mathbf{\widetilde{\Sigma}}$:
\begin{equation}
\mathbf{\widetilde{\Sigma}} = 
\tens{M} \mathbf{\widehat{\Sigma}} \tens{M}^\intercal.
\label{eqn:cov_transform}
\end{equation}
We obtained the spin-2 mixing matrix \tens{M} for this calculation using \texttt{NaMaster}.

\subsection{Connected non-Gaussian covariance}

We also calculated the connected non-Gaussian covariance component using \texttt{CosmoLike} (referred to in \citealt{Krause2017CosmoLike} as the `non-Gaussian covariance in the absence of survey window effects'), using the same adapted version of the \texttt{CosmoCov} package that we make available at the URL provided in \cref{sec:ssc}. 

\texttt{CosmoLike} calculates the connected non-Gaussian covariance as the projected matter trispectrum. The trispectrum is calculated using a halo model, with the details given in eqs. A3--A6 of \citet{Krause2017CosmoLike} \citep[see also][]{Cooray2002, Takada2009}. We find this to be suitably accurate in \cref{sec:cov_vs_sims}, and we also find in \cref{sec:importance} that the contribution from the connected non-Gaussian component to the total parameter posterior error is no more than 10--20 per cent. As with the super-sample covariance component, we multiplied the connected non-Gaussian covariance matrix by the mixing matrix using \cref{eqn:cov_transform}.

The calculation of the connected non-Gaussian covariance component is much slower than the other two components. 
As a result, in \cref{sec:params} we use an approximation to directly estimate the connected non-Gaussian covariance of power spectrum bandpowers (i.e. power spectra that have been binned in multipole space), which we subsequently use to obtain mock parameter constraints. The approximation is described in that section.
However, the results provided in Sects. \ref{sec:sims} and \ref{sec:rel_sizes} are obtained using the full (i.e. per-multipole) connected non-Gaussian covariance matrix for a single redshift bin. This took 36 days to calculate on 55 CPUs for 32 million elements.\footnote{This number comes from a data vector running from $\ell_\text{min} = 2$ to $\ell_\text{min} = 8000$, which has a length of $n = 8000 - 2 + 1 = 7999$, leading to a number of unique covariance elements equal to $n \left( n + 1 \right) / 2 = 31\,996\,000$.} By contrast, the equivalent Gaussian and super-sample covariance matrices each took around an hour to calculate on 12 CPUs.

\section{Comparison to simulations}
\label{sec:sims}

\begin{figure*}
\includegraphics[width=\textwidth]{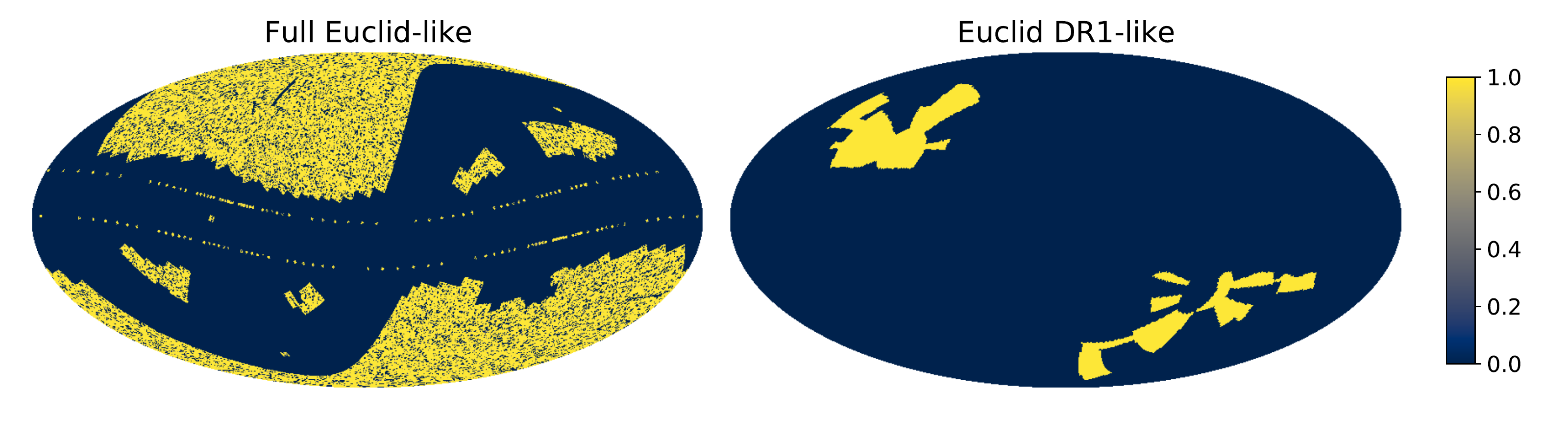}
\caption{Full \Euclid-like and \Euclid DR1-like masks, which we use in Sects. \ref{sec:sims} and \ref{sec:importance} to quantify the effects of different masks on the power spectrum covariance.}
\label{fig:masks}
\end{figure*}

\begin{figure*}
\includegraphics[width=.869\textwidth]{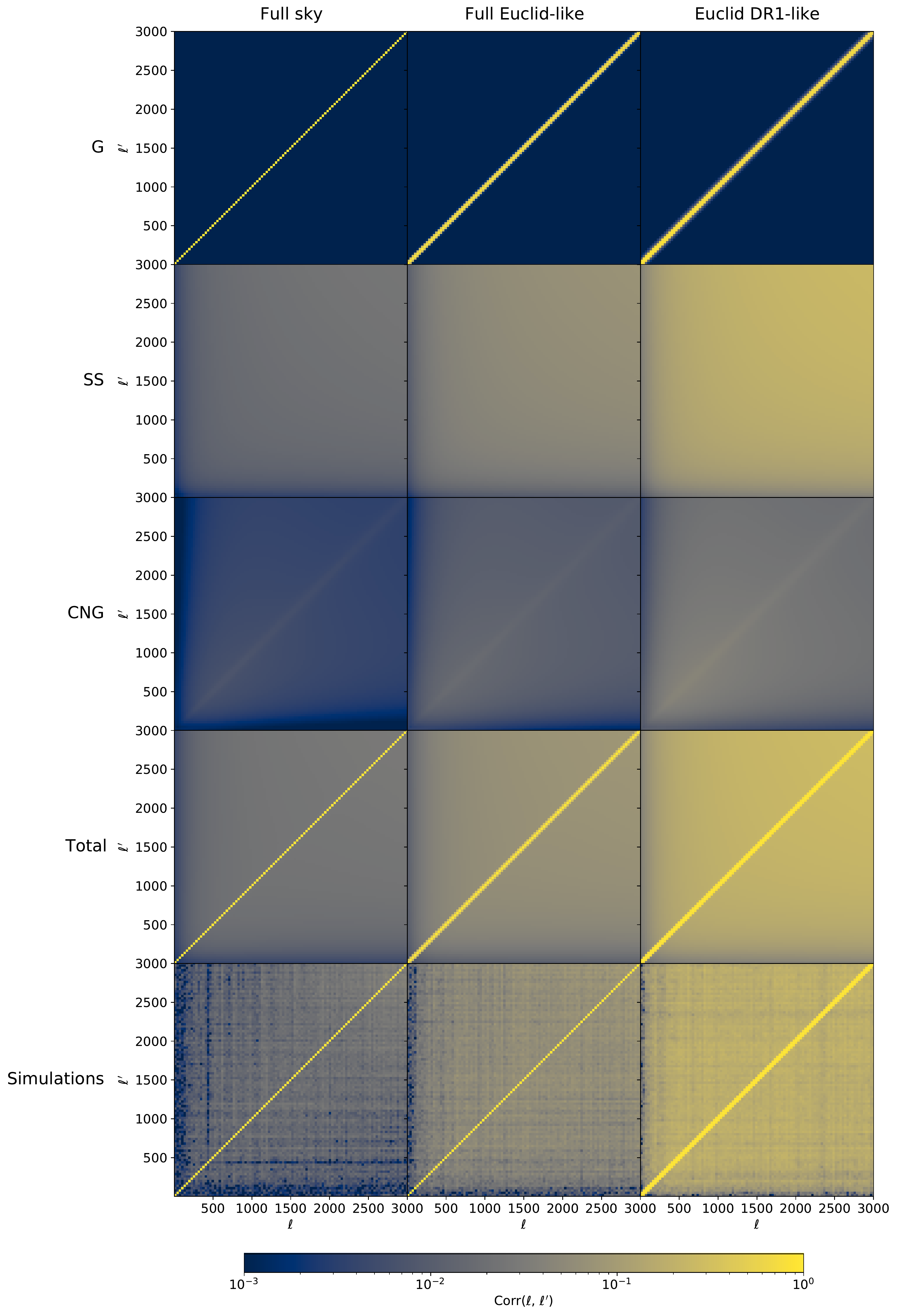} 
\caption{Correlation matrices for the simulated covariance compared to the total theoretical covariance for each mask and for the individual components of the theory covariance: Gaussian (G), super-sample (SS), and connected non-Gaussian (CNG). The covariance shown here is for the shear $E$-mode power spectrum in the lowest redshift bin, without shape noise.}
\label{fig:cov_mats}
\end{figure*}

\begin{figure*}
\includegraphics[width=\textwidth]{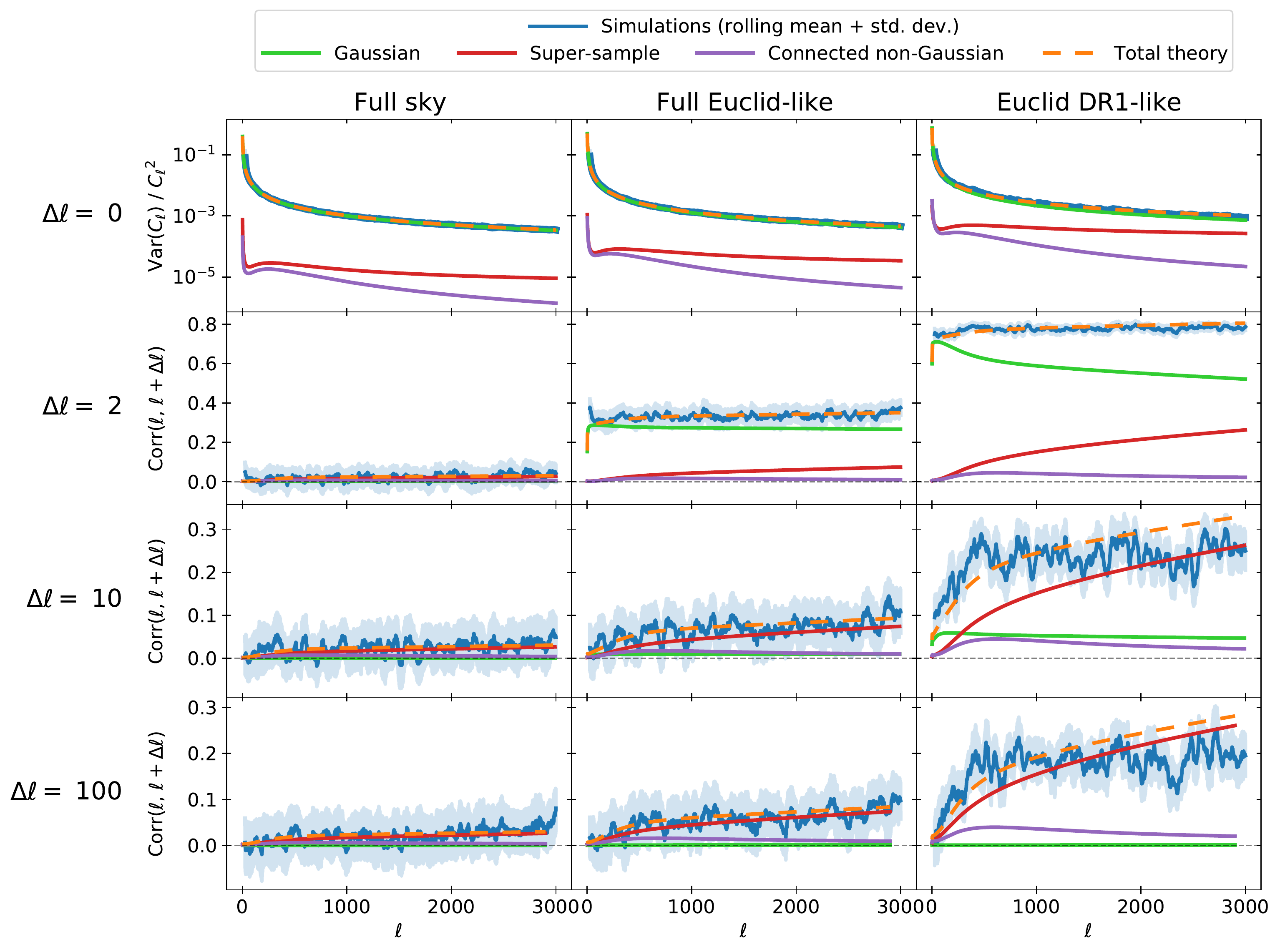}
\caption{Comparison between the covariance predicted by theory and measured from simulations, for the three masks. The top row shows the variance divided by the power spectrum squared, and the lower three rows show correlation. In all panels the simulated line is a rolling average over 50 $\ell$s, and the shaded region is the standard deviation over this range. The covariance shown here is for the shear $E$-mode power spectrum in the lowest redshift bin, without shape noise.}
\label{fig:cov_diags}
\end{figure*}

\begin{figure*}
\includegraphics[width=\textwidth]{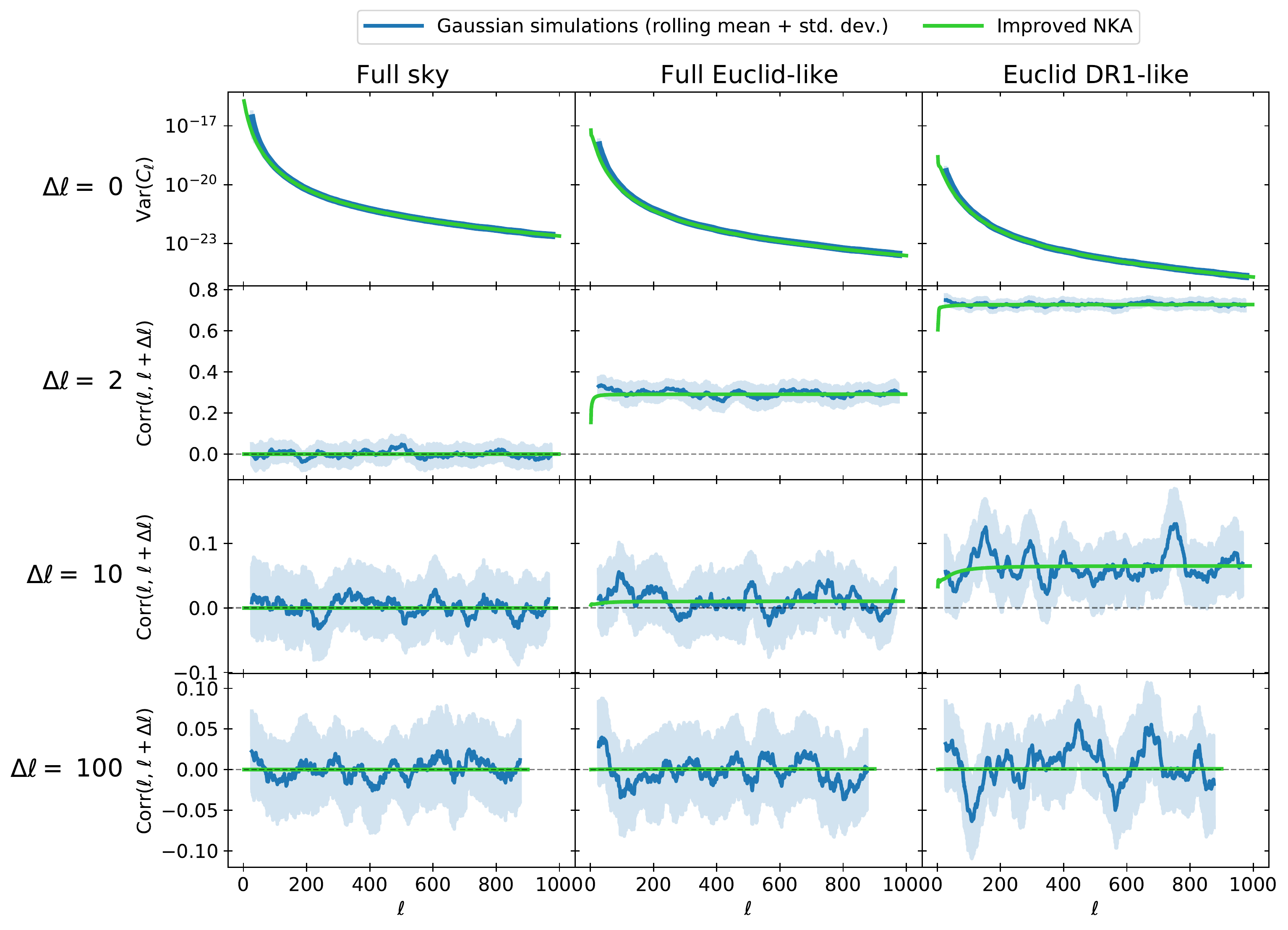}
\caption{Comparison between covariance measured from Gaussian field simulations and predicted using the improved NKA method, for the three masks. In all panels the simulated line is a rolling average over 50 $\ell$s, and the shaded region is the standard deviation over this range. No shape noise is included.}
\label{fig:cov_diags_gauss}
\end{figure*}

\subsection{Method}

We used the publicly available\footnote{\href{http://cosmo.phys.hirosaki-u.ac.jp/takahasi/allsky_raytracing}{http://cosmo.phys.hirosaki-u.ac.jp/takahasi/allsky\_raytracing}} full-sky simulated spin-2 shear maps of \citet{Takahashi2017}, which were produced by ray tracing through cosmological N-body simulations. The simulations use a maximum box size of $6300 \, h^{-1} {\rm Mpc}$. These simulated maps are quite versatile, not only because they cover the full sky, but also because they describe the underlying shear field with no shape noise or shot noise (which could be added if required, but we chose not to here). A total of 108 realisations were performed, which is a relatively small number considering that very large numbers of realisations may be required for full convergence of a simulated covariance \citep{Blot2016}, but we find that a useful comparison between theory and simulations is still possible. In addition, finite box simulations will necessarily underestimate the super-sample covariance \citep{Hamilton2006, Li2014}, but we do not detect a significant deficit. We refer the reader to \citet{Takahashi2017} for further details and validation of the simulations, and we note that density maps, halo catalogues and lensed cosmic microwave background maps are also made available at the same URL.

The shear maps are provided at 38 redshift slices from $z = 0.05$ to 5.3. 
For each realisation, we combined these into five redshift bins, following a Gaussian redshift distribution centred at $z = 0.65$, 0.95, 1.25, 1.55, 1.85 with a standard deviation of 0.3. 
The combined shear map for each redshift bin is formed as a weighted average over all 38 slices, with the weights given by the probability density of a Gaussian distribution with the appropriate mean and standard deviation. We comment on this choice of redshift distribution below, in \cref{sec:redshift}.
We then took three copies: one full-sky with no mask, one with a full \Euclid-like mask including the survey footprint and a bright star mask ($f_\text{sky} = 0.31$), and one with a \Euclid DR1-like footprint but no bright star mask ($f_\text{sky} = 0.06$). The full \Euclid-like and \Euclid DR1-like masks are shown in \cref{fig:masks}. These masks approximate the coverage of the Euclid Wide Survey at different stages but do not exactly correspond to what will be observed, which is described in \cite{Scaramella2021}. We also assume that the masks are uncorrelated with the signal, which may not be the case in practice \citep[e.g.][]{Fabbian2021}. Finally, we used the \texttt{healpy}\footnote{\href{https://github.com/healpy/healpy}{https://github.com/healpy/healpy}} \citep{Zonca2019} interface to the \texttt{HEALPix}\footnote{\href{http://healpix.sourceforge.net}{http://healpix.sourceforge.net}} \citep{Gorski2005} software to measure the spin-2 shear power spectra for each realisation. The comparisons shown in this section are for the $E$-mode auto-power in the lowest redshift bin.

For the theoretical covariance components described in \cref{sec:theory}, we used the same cosmology and redshift distribution as the simulations. We used $\ell_\text{max} = 8000$ in intermediate calculations to fully account for all relevant mode coupling, but we limited the comparison to $\ell \leq 3000$ because the $n_\text{side} = 4096$ maps that we used experience distortion from limited angular resolution above this point, as documented in \citet{Takahashi2017}. We note that $n_\text{side} = 8192$ maps are also available, so the $\ell$ range could in principle be extended, albeit with significantly increased computational requirements.

\subsubsection{Choice of redshift distribution}
\label{sec:redshift}

The choice of redshift distribution used here -- five Gaussian bins, centred on $z = 0.65$, 0.95, 1.25, 1.55, 1.85 with a standard deviation of 0.3 -- was made for simplicity,  with the relatively low number of redshift bins (five) also chosen for computational efficiency. Since we have the freedom to enforce agreement between the redshift distributions in the simulations and theory, there is little additional value in choosing a more complicated, more realistic distribution. Future cosmological analyses of real \Euclid{} data are likely to use a larger number of bins with less overlap than is used here \citep[e.g.][]{Pocino2021}. There is no reason that this will affect the results presented in this paper, although we note that marginalising over nuisance parameters describing photometric redshift uncertainties will reduce the importance of all cosmological contributions to the covariance.

\subsection{Results}
\label{sec:cov_vs_sims}

In \cref{fig:cov_mats} we show correlation matrices for the simulated covariance compared to the total theoretical covariance for each mask, as well as the individual components of the theory covariance. There appears to be good agreement between the simulated and total theoretical correlation matrices for all three masks. We discuss the relative contributions from the three covariance components in \cref{sec:importance}.

In \cref{fig:cov_diags} we show a detailed comparison of certain diagonals of the covariance matrix. For the main diagonal ($\Delta \ell = 0$) we show the variance divided by the square of the power spectrum, to remove any effects coming from disagreement in the power spectrum between the simulations and theory, which is not our focus here. For the off-diagonals ($\Delta \ell =$ 2, 10, 100) we show the correlation. In all panels the simulated line is a rolling average over 50 $\ell$s and the shaded region is the standard deviation over this range. For the full-sky and full \Euclid-like masks, we find excellent agreement between the theory and simulations. For the more extreme \Euclid DR1-like mask there is a slightly worse, but still generally good, level of agreement. In particular, the super-sample covariance component is clearly correctly increasing with the severity of the sky cut to match the additional correlation found in the simulations. We discuss the relative sizes and importance of the three theory contributions further in \cref{sec:importance}. We conclude from Figs. \ref{fig:cov_mats} and \ref{fig:cov_diags} that \texttt{CosmoLike}'s non-Gaussian covariance calculations appear to be suitably accurate, to the degree that can be assessed using these simulations.

We also repeat this comparison for purely Gaussian fields, which we simulated using \texttt{healpy}. The results are shown in \cref{fig:cov_diags_gauss}, where we find an excellent level of agreement between the Gaussian field simulations and the prediction from the improved NKA, which is a significant improvement over the standard NKA (not shown).

\section{Importance of covariance components and dependence on mask}
\label{sec:importance}

In this section we discuss the size and importance of the different components of the cosmic shear pseudo-$C_\ell$ covariance, and how these properties depend on the mask.

\subsection{Relative sizes of components}
\label{sec:rel_sizes}

\begin{figure*}
\includegraphics[width=\textwidth]{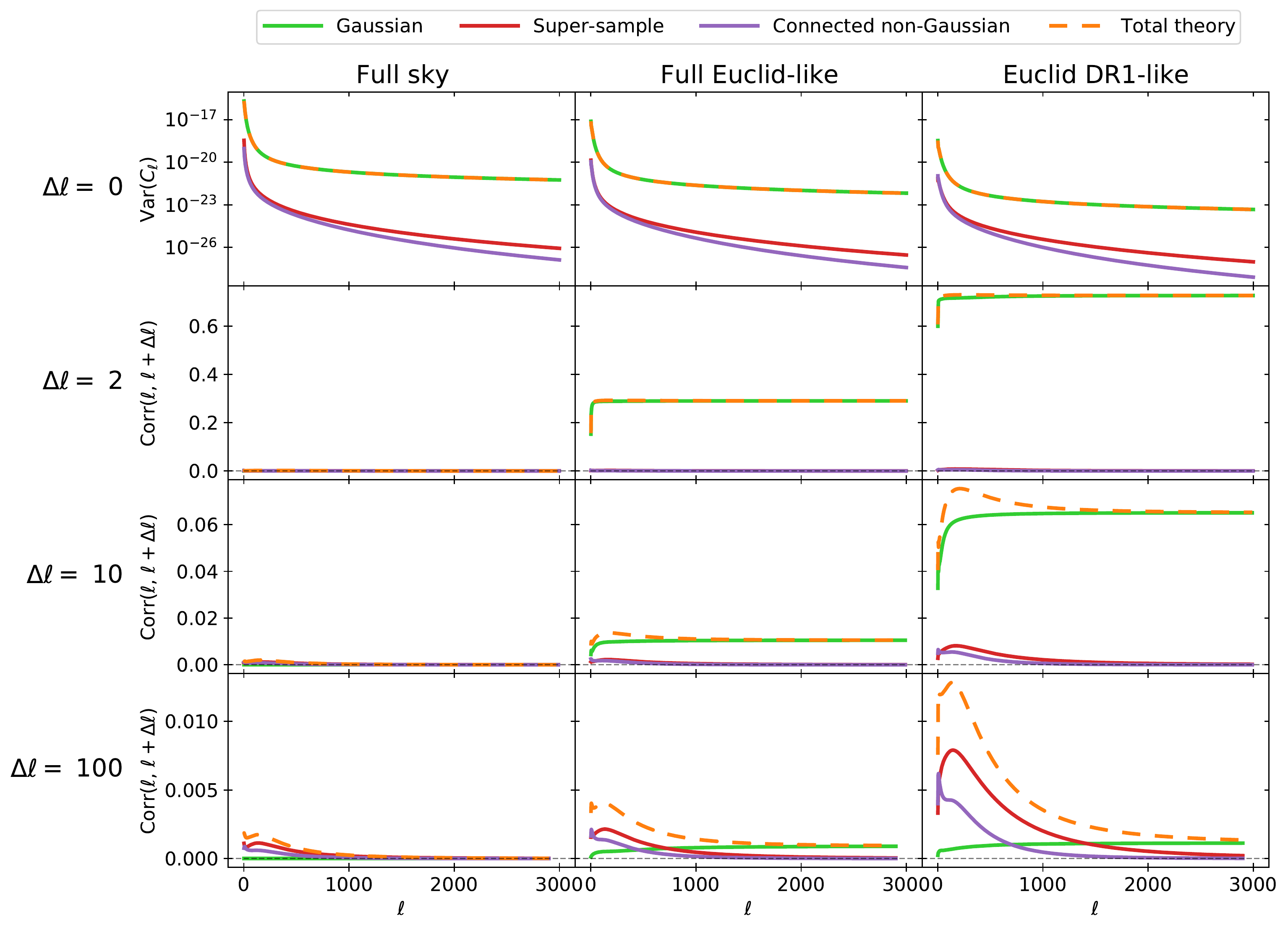}
\caption{Comparison between contributions to the theoretical covariance for the three masks, with shape noise included 
following \cref{eqn:nl}. The top row shows the variance, and the lower three rows show correlation.}
\label{fig:cov_diags_withnoise}
\end{figure*}

\subsubsection{Without shape noise}

We first discuss the case without shape noise, which has already been shown in Figs. \ref{fig:cov_mats} and \ref{fig:cov_diags}. We find from the full correlation matrices plotted in \cref{fig:cov_mats} that the main diagonal of the matrix is dominated by the Gaussian component, which is purely diagonal in the full-sky case and visibly broadens slightly as the sky cut is increased. The super-sample covariance is the dominant off-diagonal component, particularly at higher $\ell$ but extending down visibly even to $\ell < 500$ in the case of the most extreme \Euclid DR1-like mask. The connected non-Gaussian contribution is barely visible on the colour scale, other than for the \Euclid DR1-like mask at low $\ell$.

A more detailed comparison of the relative sizes of the different components is possible with the selected diagonals shown in \cref{fig:cov_diags}. We find again that the main diagonal ($\Delta \ell = 0$) is dominated by the Gaussian component, but the extent to which this is the case is reduced as the sky cut is increased, as the contribution increases from both non-Gaussian components. Moving away from the main diagonal, at $\Delta \ell = 10$ the Gaussian is still the largest component (except on the full sky, where its contribution is purely diagonal), but by $\Delta \ell = 100$ the super-sample component is dominant and increasing towards higher $\ell$. It is clear that the super-sample covariance contribution increases with a more severe sky cut, though notably it is still visibly non-zero even for full-sky observations. The connected non-Gaussian component is the subdominant non-Gaussian contribution at all values of $\ell$ and $\Delta \ell$ for all masks.

\subsubsection{With shape noise}

In \cref{fig:cov_diags_withnoise} we show an equivalent comparison of the sizes of the theoretical covariance components with shape noise included. We assumed Gaussian shape noise and included it as a contribution to the power spectrum,
\begin{align}
C_\ell &\rightarrow C_\ell + N_\ell; \\
N_\ell &= \frac{{\sigma_\varepsilon}^2}{N_i}, 
\label{eqn:nl}
\end{align}
where $\sigma_\varepsilon$ is the intrinsic galaxy shape dispersion per component and $N_i$ is the number of galaxies per steradian per redshift bin. We assumed a \Euclid-like galaxy number density of 30\,$\text{arcmin}^{-2}$ equally split over five redshift bins, and a value of $\sigma_\varepsilon = 0.3$.

With shape noise included, we find quite different behaviour to the no-noise case. The Gaussian-dominated main diagonal is substantially increased, especially at higher $\ell$, resulting in non-Gaussian off-diagonal correlations being significantly suppressed. The result is that the Gaussian component is dominant at all $\ell$ as far away from the main diagonal as $\Delta \ell = 10$. By $\Delta \ell = 100$, the Gaussian component is no longer dominant at lower $\ell$, but continues to account for the largest contribution at higher $\ell$: above $\ell \sim 1000$ for the full \Euclid-like mask and $\ell \sim 1500$ for the \Euclid DR1-like mask. This suggests that once shape noise is included, the Gaussian component is more important than the no-noise results in \cref{fig:cov_diags} suggest.

\subsection{Importance for parameter constraints}
\label{sec:params}

\begin{figure*}
\includegraphics[width=\textwidth]{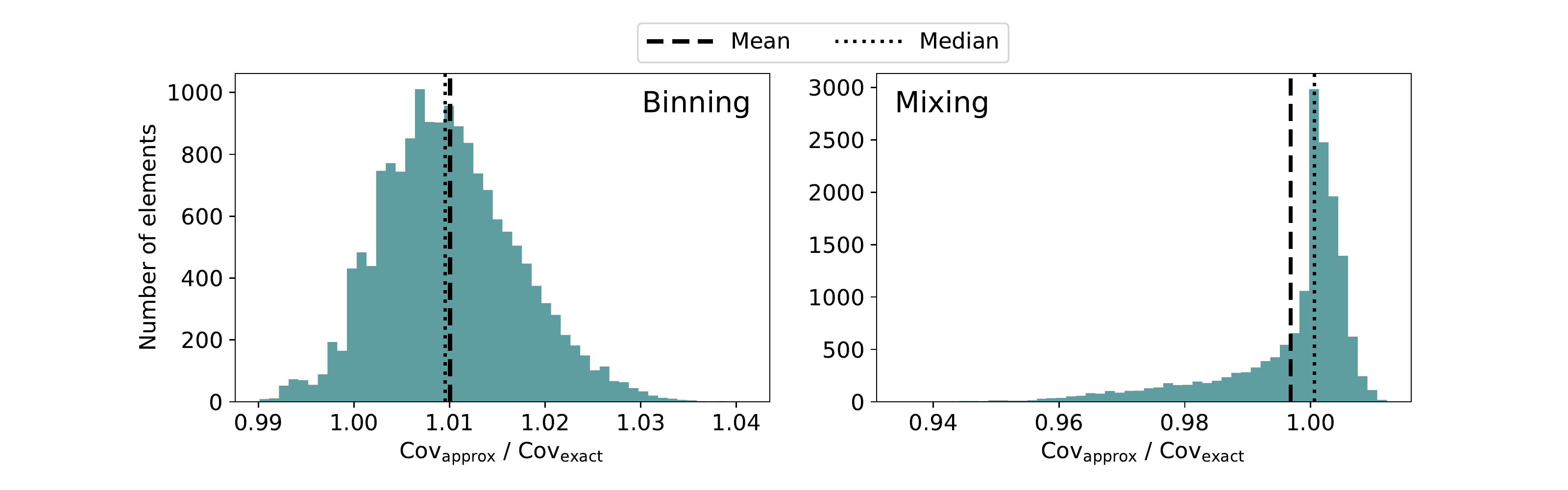}
\caption{Validation of the connected non-Gaussian approximation used to obtain the mock parameter constraints in \cref{sec:params}, which is described in \cref{sec:cng_approx}. Histograms of the ratio of the approximate to exact covariance are shown, for the `binning' (left) and `mixing' (right) steps, for all elements of the bandpower covariance matrix across all redshift bins, measured using the super-sample covariance. The results in all other sections are obtained using the connected non-Gaussian component, calculated in full.}
\label{fig:cng_approx}
\end{figure*}

\begin{figure*}
\includegraphics[width=\textwidth]{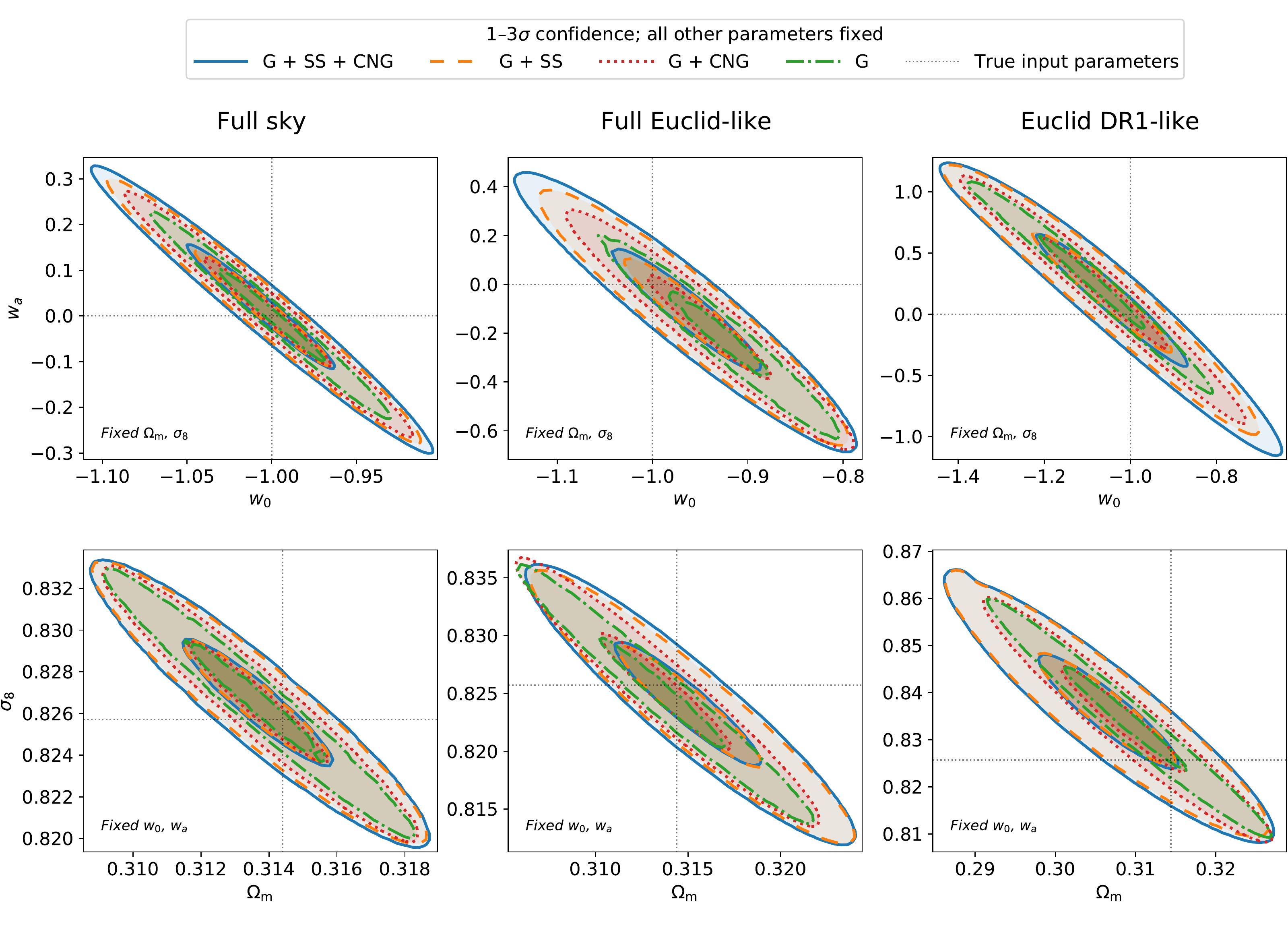}
\caption{Two-parameter constraints for different masks and different combinations of covariance contributions: Gaussian (G), super-sample (SS), and connected non-Gaussian (CNG). Shape noise is included, following \cref{eqn:nl}. In each panel, all parameters other than the two shown are held fixed. Only the 1 and 3\,$\sigma$ confidence regions are marked, which respectively contain the highest 68.3 and 99.7 per cent of the posterior probability mass. The relative areas of each $3\sigma$ confidence region are listed in \cref{tbl:rel_areas}. We note that the axis ranges differ between panels.}
\label{fig:2d_constraints}
\end{figure*}

While the relative size of the different covariance components studied in \cref{sec:rel_sizes} offers interesting insight into their behaviour, it says relatively little about the actual importance of each component. In particular, it is unclear to what extent the dominance of the Gaussian component on and close to the main diagonal is offset by its sub-dominance farther away from the main diagonal. Here, to gain more insight into this, we produce mock parameter constraints. Shape noise is included in this section, following \cref{eqn:nl}.

Here we used five redshift bins, including all auto- and cross-spectra ($E$-modes only), giving 15 power spectra in total. We used scales up to $\ell_\text{max} = 5000$. The full data vector for this setup would have $n =$ 75\,000 elements ($15 \times 5000$), which gives $n \left( n + 1 \right) / 2 =$ 2.8\,billion unique covariance elements. Due to the time needed to evaluate the projected matter trispectrum, it would be unfeasible to calculate the connected non-Gaussian contribution in full. As a result, we elected to bin into 12 logarithmically spaced bandpowers and use an approximation to obtain the connected non-Gaussian bandpower component from a more modest number of per-$\ell$ covariance calculations. We describe this approximation in \cref{sec:cng_approx}. The Gaussian and super-sample covariance components were calculated in full, using scales up to $\ell_\text{max} = 8000$ for intermediate calculations, before being binned into bandpowers as
\begin{equation}
P_b = \sum_\ell \tens{P}_{b \ell} C_\ell,
\label{eqn:cl_to_bp}
\end{equation}
where \tens{P} is the bandpower binning matrix whose elements are given by
\begin{equation}
\tens{P}_{b \ell} =
\begin{cases}
\dfrac{\ell \left( \ell + 1 \right)}{2 \pi}
\left[ \ell_\text{min}^{b + 1} - \ell_\text{min}^b \right]^{-1}
& \text{for } \ell_\text{min}^b \leq \ell < \ell_\text{min}^{b + 1}; \\
0 & \text{otherwise,}
\end{cases}
\label{eqn:pbl}
\end{equation}
where $\ell_\text{min}^b$ is the lower edge of bin $b$.

We obtained a mock observation by sampling from a Gaussian likelihood with the total covariance. The input mean was the fiducial theory power spectra, plus noise for auto-spectra given by \cref{eqn:nl}, mixed using the mixing matrix obtained using \texttt{NaMaster}, then binned following \cref{eqn:cl_to_bp}. This random sampling process replicates the randomness of cosmic variance that is present in a real observation, and means that -- as with real data -- the resulting posterior distributions are not centred on the `true' input parameters. 
We verified that bandpowers measured from the \citet{Takahashi2017} simulations used in \cref{sec:sims} are no more non-Gaussian than those measured from Gaussian field simulations, and therefore since a Gaussian likelihood was shown to be sufficiently accurate for Gaussian fields in \citet{Upham2021}, it is a suitable choice here.
To obtain parameter constraints, we iterated over two-parameter grids produced using \texttt{CosmoSIS}\footnote{\href{https://bitbucket.org/joezuntz/cosmosis}{https://bitbucket.org/joezuntz/cosmosis}} \citep{Zuntz2015}, following the pipeline described in sect. 2.2.1 of \citet{Upham2021}.\footnote{The pipeline includes the \texttt{CAMB} \citep{Lewis2000, Howlett2012} and \texttt{Halofit--Takahashi} \citep{Smith2003, Takahashi2012} modules.} All other parameters were held fixed. At each point in parameter space, we compared theory bandpowers -- calculated in the same way as the input mean to the observation described above -- to the observed bandpowers using a Gaussian likelihood with different combinations of covariance components. All combinations necessarily include the Gaussian component, since this on its own is a valid positive definite covariance matrix, unlike the super-sample and connected non-Gaussian components. 

\subsubsection{Connected non-Gaussian approximation}
\label{sec:cng_approx}

As noted above, it is impractical to calculate the connected non-Gaussian component for all 2.8 billion unique elements of the full covariance matrix. Instead we used an approximation to directly obtain its contribution to the bandpower covariance. 
This approximation was designed to mimic two effects: the mixing of power by the survey mask, and the binning of individual multipoles into bandpowers. In our analysis, both of these processes are cosmology-independent: each shear field uses the same mask, and all power spectra use the same binning scheme. As a result, both processes should have approximately the same effect on every power spectrum, and consequently also every covariance block. Therefore, the approximation that we make used two sets of weights -- one to mimic binning and the other to mimic mixing -- which were calibrated for the covariance of the shear auto-power spectrum in the lowest redshift bin and then applied to all further blocks.

First, we evaluated in full the connected non-Gaussian component for a single covariance element per bandpower pair, for all combinations of power spectra. This was chosen to be for the weighted average $\ell$ in each bandpower, with the weights given by $\tens{P}_{b \ell}$ (\autoref{eqn:pbl}), rounded to the nearest integer. This vastly reduced the number of projected trispectrum calculations, to 16\,290.\footnote{This number comes from a reduced data vector of 12 bandpowers and 15 power spectra, giving a data vector of length $n = 12 \times 15 = 180$ and a number of unique covariance elements of $n \left( n + 1 \right) / 2 = 16\,290$.} We then re-weighted the result using the weights calibrated using the covariance of the shear auto-power spectrum in the lowest redshift bin, which was calculated in full for the previous sections.

The weighting can be understood as a two-step process. First, we applied a `binning' weighting, designed to mimic the effect of taking the full unbinned covariance matrix and binning it into bandpowers. Then we applied a `mixing' weighting, designed to mimic the effect of the mixing matrix. In both cases, the weights were obtained by carrying out the process in full for the shear auto-power spectrum in the lowest redshift bin.

This can be illustrated using equations as follows. For the first block (covariance of shear auto-power in the lowest redshift bin), the following procedure was used to transform the full unbinned covariance $\tens{Cov}_\text{unbinned}$ into a final binned and mixed block $\tens{Cov}_\text{mixed}$, via a binned and unmixed stage $\tens{Cov}_\text{binned}$:
\begin{align}
\tens{Cov}_\text{binned} &= 
\tens{P} ~ \tens{Cov}_\text{unbinned} ~ 
\tens{P}^\intercal; \\
\tens{Cov}_\text{mixed} &= 
\tens{M} ~ \tens{Cov}_\text{binned} ~ 
\tens{M}^\intercal,
\end{align}
where \tens{P} and \tens{M} are the bandpower binning and pseudo-$C_\ell$ mixing matrices, respectively. Elements were selected from $\tens{Cov}_\text{unbinned}$ corresponding to the weighted average $\ell$ within each bandpower to give $\tens{Cov}_\text{sampled}$. The matrices of binning weights $\tens{w}_\text{bin}$ and mixing weights $\tens{w}_\text{mix}$ were then calculated as
\begin{align}
\tens{w}_\text{bin} = 
\frac{\tens{Cov}_\text{binned}}
{\tens{Cov}_\text{sampled}}
\qquad&\text{(elementwise);} \\
\tens{w}_\text{mix} = 
\frac{\tens{Cov}_\text{mixed}}
{\tens{Cov}_\text{binned}}
\qquad&\text{(elementwise).}
\end{align}
Finally, the sampled covariance blocks  $\tens{Cov}_\text{sampled}$ were calculated for every block in the full covariance matrix, and transformed to give approximate binned and mixed covariance blocks as
\begin{align}
\tens{Cov}_\text{binned\_approx} &= 
\tens{w}_\text{bin} *
\tens{Cov}_\text{sampled}
~&\text{(elementwise);} \\
\tens{Cov}_\text{mixed\_approx} &= 
\tens{w}_\text{mix} *
\tens{Cov}_\text{binned\_approx}
~&\text{(elementwise).}
\end{align}
While this two-step weighting process could be equivalently formulated as a single step, we chose to separate the effect of the binning and mixing approximations for additional insight into their respective effects.

We were able to validate these approximations by carrying out an equivalent process for the super-sample covariance matrix and comparing the results to those obtained using the full correct treatment. Histograms of the ratios between the approximate and exact covariance for each step, for all elements of the bandpower covariance across all redshift bins, are shown in \cref{fig:cng_approx}. We find that each step introduces a bias of order per cent on average (although conveniently in opposite directions) with a spread of a few per cent. We find this to be sufficiently accurate for our purposes here, especially considering that the connected non-Gaussian component is the smallest of the three, but nevertheless this small potential error should be borne in mind when interpreting the results. Since the super-sample and connected non-Gaussian covariance contributions were shown to be similarly smooth in \cref{sec:sims}, the fact that this approximation works well for the former suggests that it should too for the latter.

\subsubsection{Results}

\begin{table*}
\caption{Relative areas of $3\sigma$ confidence regions in \cref{fig:2d_constraints}.}
\label{tbl:rel_areas}
\begin{tabular}{llrrrr}
\hline
\multirow{2}{*}{Parameters} & \multirow{2}{*}{Mask} & \multicolumn{4}{c}{Relative area of $3\sigma$ confidence region (\%)} \\
& & G + SS + CNG & G + SS & G + CNG & G \\ \hline \hline
\multirow{3}{*}{($w_0$, $w_a$)}
& Full sky & 100 & 82 & 63 & 38 \\
& Full \Euclid-like mask & 100 & 84 & 61 & 36 \\
& \Euclid DR1-like mask & 100 & 85 & 51 & 30 \\ \hline
\multirow{3}{*}{($\Omega_\text{m}$, $\sigma_8$)}
& Full sky & 100 & 90 & 70 & 49 \\
& Full \Euclid-like mask & 100 & 90 & 66 & 45 \\
& \Euclid DR1-like mask & 100 & 92 & 54 & 37 \\ \hline
\end{tabular}
\tablefoot{The table shows the relative area of the $3\sigma$ confidence region in \cref{fig:2d_constraints} for each combination of parameters and mask and for different combinations of covariance components: Gaussian (G), super-sample (SS), and connected non-Gaussian (CNG). The relative $1\sigma$ areas are similar to the $3\sigma$ areas.}
\end{table*}

Two-parameter constraints for different combinations of covariance components are shown in \cref{fig:2d_constraints}. 
The top row shows dark energy equation of state parameters ($w_0$, $w_a$), where $w \left( a \right) = w_0 + w_a \left( 1 - a \right)$. 
The bottom row shows the matter density $\Omega_\text{m}$ and the amplitude of the matter power spectrum at $z = 0$ on the scale of $8\,h^{-1}\,\text{Mpc}$, $\sigma_8$.
The three columns are for the three different masks. In each panel, all parameters other than the two shown are held fixed. Only the 1 and 3\,$\sigma$ confidence regions are marked, which respectively contain the highest 68.3 and 99.7 per cent of the posterior probability mass. We list the relative areas of the $3\sigma$ confidence region for each combination of parameters and mask in \cref{tbl:rel_areas}.

We find that the Gaussian contribution (G) alone only covers 30--38 per cent of the full $3\sigma$ region for ($w_0$, $w_a$) and 37--49 per cent for ($\Omega_\text{m}$, $\sigma_8$). The Gaussian and connected non-Gaussian components combined (G + CNG) cover 51--63 and 54--70 per cent for ($w_0$, $w_a$) and ($\Omega_\text{m}$, $\sigma_8$), respectively, while the Gaussian and super-sample components combined (G + SS) cover 82--84 and 90--92 per cent. These results are broadly in line with the single-parameter error bar ratios obtained in \citet{Barreira2018b}.

There is some amount of apparent mask dependence: we find that as the sky cut is increased, the relative area of G + SS sees a very small increase (by 2--3 per cent) whereas G + CNG and G see larger decreases (by 16--22 and 8--12 per cent, respectively). This is consistent with the expectation that super-sample covariance should become more important as the sky is cut further, since this excludes more modes from the survey.

We also find some small shifts in the posterior means between the different composite covariance results for a given mask, despite the fact that the mean of the Gaussian likelihood used in each case is identical and only the covariance differs. This demonstrates how an incorrect covariance leads to an incorrect weighting of the random scatter present in the data due to cosmic variance, and therefore to posterior constraints having not only the wrong size but an erroneous position too, although this is a small effect.

\subsubsection{Effect of marginalisation over additional parameters}

\begin{figure*}
\includegraphics[width=\textwidth]{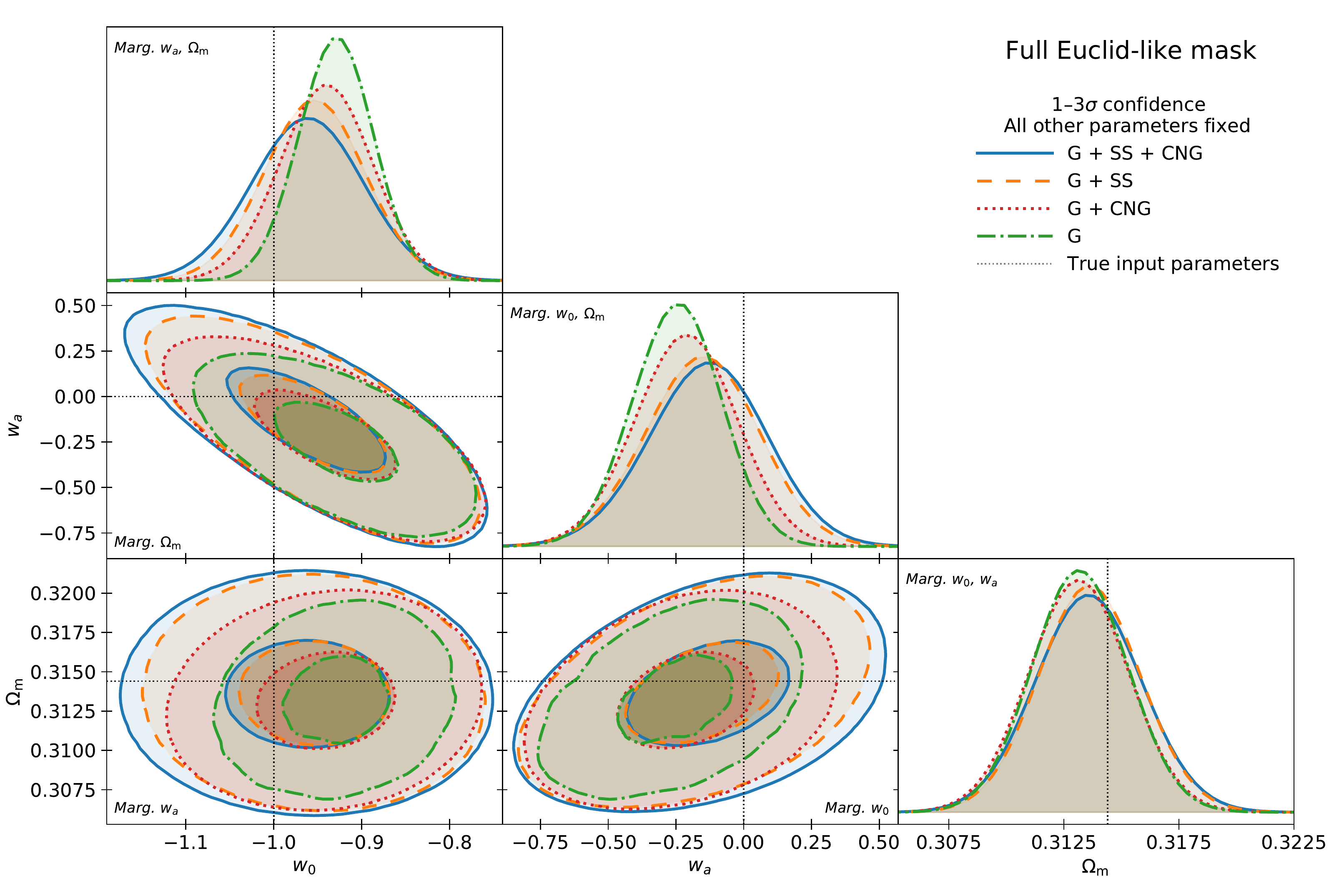}
\caption{Two- and one-parameter marginalised constraints obtained from a joint three-parameter analysis of ($w_0$, $w_a$, $\Omega_\text{m}$) for the full \Euclid-like mask, including different combinations of covariance contributions: Gaussian (G), super-sample (SS), and connected non-Gaussian (CNG). Shape noise is included, following \cref{eqn:nl}. The constraints in each panel have been obtained by marginalising over one or two parameters in the joint three-parameter posterior; for example, the panel marked `Marg. $w_a$, $\Omega_\text{m}$' has been marginalised over $w_a$ and $\Omega_\text{m}$. All other parameters are held fixed. Only the 1 and 3\,$\sigma$ confidence regions are marked.}
\label{fig:3d_constraints}
\end{figure*}

\begin{table*}
\caption{Impact of marginalisation on $3\sigma$ confidence regions.}
\label{tbl:marg}
\begin{tabular}{llrrrr}
\hline
\multirow{2}{*}{Parameter(s)} & \multirow{2}{*}{Marginalised over} & \multicolumn{4}{c}{Relative area or width of $3 \sigma$ confidence region (\%)} \\
& & G + SS + CNG & G + SS & G + CNG & G \\ \hline\hline
\multirow{2}{*}{($w_0$, $w_a$)}
& - & 100 & 84 & 61 & 36 \\
& $\Omega_\text{m}$ & 100 & 89 & 84 & 69 \\ \hline
\multirow{2}{*}{$w_0$}
& $w_a$ & 100 & 91 & 88 & 72 \\
& ($w_a$, $\Omega_\text{m}$) & 100 & 90 & 83 & 69 \\ \hline
\multirow{2}{*}{$w_a$}
& $w_0$ & 100 & 88 & 84 & 70 \\
& ($w_0$, $\Omega_\text{m}$) & 100 & 96 & 85 & 74 \\ \hline
\end{tabular}
\tablefoot{The table shows the relative areas (for two-parameter constraints) and widths (for one-parameter constraints) of the $3\sigma$ confidence regions obtained using different combinations of covariance contributions, Gaussian (G), super-sample (SS), and connected non-Gaussian (CNG), for the full \Euclid-like mask. Each row contains two sub-rows: the top sub-row is based on a two-parameter fit, which is marginalised over zero or one parameters; the bottom sub-row is based on a three-parameter fit, which is marginalised over one or two parameters.The relative $1\sigma$ areas are similar to the $3\sigma$ areas.}
\end{table*}

We are motivated to explore the impact of marginalisation over additional parameters on the relative importance of each covariance component by the results of \citet{Barreira2018b}. In that paper, the authors produced mock parameter constraints with different covariance contributions included, both for a single parameter at a time with all others fixed and for five parameters with all five allowed to vary simultaneously. For the latter case, they display marginalised two-parameter constraints \citep[their Fig. 3]{Barreira2018b}. For ($w_0$, $w_a$) in particular (and to a lesser extent with some other parameter pairs), the $2\sigma$ confidence region obtained with the Gaussian covariance alone appears roughly the same as that for the total covariance, to within the sampling noise. This is in contrast to their single-parameter constraints, for which the Gaussian covariance only produced $\sim$ 50 per cent of the $1\sigma$ uncertainty on $w_0$ obtained using the full covariance (see Fig. 2 of \citealt{Barreira2018b}). This raises the question of whether marginalisation might reduce the differences between the constraints obtained using different combinations of covariance components.

We explore this question by performing a three-parameter likelihood analysis using the full \Euclid-like mask over ($w_0$, $w_a$, $\Omega_\text{m}$), with all other parameters still held fixed. We show one- and two-parameter marginalised constraints in \cref{fig:3d_constraints}. In \cref{tbl:marg} we compare relative areas and widths of one- and two-parameter $3\sigma$ confidence regions before and after marginalisation over a third parameter.

While the relative areas in \cref{fig:3d_constraints} appear qualitatively similar to those in \cref{fig:2d_constraints}, we find that there are in fact some substantial quantitative differences, as shown by the values in \cref{tbl:marg}. In particular, there is almost a doubling in the area of the constraints on ($w_0$, $w_a$) from the Gaussian covariance only (G) relative to the total covariance (G + SS + CNG) -- from 36 to 69 per cent -- when marginalising over $\Omega_\text{m}$ rather than holding it fixed. A similar but smaller increase is seen for the other subsets (G + SS, G + CNG) of the total covariance for the same parameters. 

For constraints on $w_a$ alone, we find that the width of the $3\sigma$ confidence region for G relative to G + SS + CNG increases slightly when marginalising over both $w_0$ and $\Omega_\text{m}$ compared to only marginalising over $w_0$, and a similar slight increase is seen for G + SS and G + CNG. However, for constraints on $w_0$, we find a small decrease in relative widths when marginalising over both $w_a$ and $\Omega_\text{m}$ rather than only $w_a$. One reason for this difference in behaviour between $w_0$ and $w_a$ may be that -- as seen in \cref{fig:3d_constraints} -- there is clearly a much stronger correlation between $w_a$ and $\Omega_\text{m}$ than between $w_0$ and $\Omega_\text{m}$. Marginalisation over a strongly correlated parameter should broaden constraints more than marginalisation over a more weakly correlated parameter (indeed, marginalisation over a truly independent parameter should have no effect at all), but it is not obvious that this should change the ratio of relative areas rather than simply broadening all constraints by the same factor. Regardless of the origin of this behaviour, it does appear to be the case that marginalisation over additional parameters -- particularly those with which the constrained parameters are correlated -- affects the relative importance of the difference covariance contributions. This is in agreement with the findings of \cite{Barreira2018b}. 

\section{Conclusions}
\label{sec:conclusions}

As the era of next-generation weak lensing surveys such as \Euclid rapidly approaches, it is increasingly important to understand the properties of all steps of an analysis pipeline, including the covariance used in the likelihood. We have described in \cref{sec:theory} how existing publicly available codes can be used in combination to calculate the full covariance matrix of cosmic shear pseudo-$C_\ell$ estimates, including the full details of an arbitrary mask. We have further shown in \cref{sec:sims} that existing simulations can be used to verify the accuracy of a theoretical covariance, and we found a high degree of agreement and consistency between theory and simulations. This agreement persists for different masks, showing that the theoretical covariance contributions correctly account for both the cut-sky mode coupling that is inherent to the pseudo-$C_\ell$ method and the non-Gaussian mode coupling, including additional cut-sky super-sample covariance. 

This is encouraging for the use of pseudo-$C_\ell$ estimators in weak lensing, whose convenience and speed make them an attractive choice of analysis framework for future surveys. However, an outstanding challenge with such estimators is the need to understand their statistical properties sufficiently well such that they can be used to deliver reliable cosmological constraints to the precision and accuracy needed by future high-precision weak lensing surveys. This challenge has now to a large degree been addressed since we now know that not only is a Gaussian likelihood sufficient \citep{Upham2021}, but a full covariance can be evaluated and validated using the methods shown in this paper. 

We have found that it is essential to include the non-Gaussian contributions to the covariance, even though cut-sky mode coupling means that the Gaussian covariance component dominates off-diagonal modes close to the main diagonal. The relative size and importance of the Gaussian component increases when including shape noise, but we have shown in \cref{sec:importance} that only including the Gaussian component in parameter inference can lead to an underestimation of uncertainties by up to 70 per cent. The dominant non-Gaussian covariance component is the super-sample covariance, but neglecting the subdominant connected non-Gaussian covariance component can still lead to uncertainty underestimation on the scale of 10--20 per cent.
In addition, neglecting some covariance contributions can lead to biases in the position of posterior parameter constraints as well as their size.
However, a real cosmological analysis will require marginalisation over many nuisance parameters, which will decrease the relative importance of all cosmological contributions to the covariance, so these values should be taken as upper limits on the importance of each component. 
Perhaps for this reason it was found in the analysis of DES Year 3 data in \citet{Friedrich2021} that the connected non-Gaussian contribution could be entirely neglected, but our results suggest that we cannot necessarily extend this conclusion to a \Euclid{}-like survey.
However, this need not be an inconvenience, since we have shown that approximations of the kind described in \cref{sec:cng_approx} can be used to obtain the connected non-Gaussian component in a manageable amount of time with a loss of accuracy of only a few per cent on an already subdominant term. Finally, we have found that marginalisation over additional cosmological parameters may have a substantial effect on the relative importance of the different covariance components. We can conclude from this that it is important to take all marginalisation into account when, for example, determining the required accuracy of theoretical results for a particular science goal or producing forecasts of parameter constraints. 
The consistency of our results with those of \citet{Barreira2018b} implies that cut-sky mode coupling has relatively little impact on the respective importance of the covariance components.

The N-body weak lensing simulations are available at \href{http://cosmo.phys.hirosaki-u.ac.jp/takahasi/allsky_raytracing}{http://cosmo.phys.hirosaki-u.ac.jp/takahasi/allsky\_raytracing} \\ and are described in \citet{Takahashi2017}.
We additionally make available
\begin{enumerate}
\item tomographic pseudo-$C_\ell$ power spectra measured from these simulations and
\item the full connected non-Gaussian covariance matrix for the lowest redshift bin
\end{enumerate}
at \href{https://doi.org/10.5281/zenodo.5163132}{https://doi.org/10.5281/zenodo.5163132}.
All other data used in this article can be quickly generated using code that we have made available at \href{https://github.com/robinupham/shear_pcl_cov}{https://github.com/robinupham/shear\_pcl\_cov}. 

\begin{acknowledgements}
We thank the internal and external referees and colleagues for helpful feedback, which has improved the manuscript.
This work would not have been possible without the contributions to the community from the developers of \texttt{NaMaster} \citep{Alonso2019, Garcia-Garcia2019}, \texttt{CosmoLike} \citep{Krause2017CosmoLike, Fang2020CosmoCov} and the simulations of \citet{Takahashi2017}.

REU acknowledges a studentship from the UK Science \& Technology Facilities Council.

\AckEC
\end{acknowledgements}

\bibliographystyle{aa}
\bibliography{refs}

\end{document}